\documentclass[11pt]{article}
\usepackage{geometry}                
\usepackage{graphicx}
\usepackage{amssymb}
\usepackage{epstopdf}
\title{Validation of Models for the Flow of Granular Media}
\author{Jeffrey Picka}
\date{}                                           
\begin{document}
\maketitle 
\begin{abstract}
Validation of models for powder flow requires that the models be stochastic and that they be fit by statistical inference. Methods from spatial and multivariate statistics can be used for model fitting and assessment. If the quality of the fitted model is not assessed, there is a significant risk the model will fail to represent the physics of powder flow. 
\end{abstract}

\section{Introduction}

A mathematical model for powder flow must summarize what
is known about the physics of the powder and must allow useful
predictions of the physical behaviour of powders to be made. The models must be
stochastic in order to represent what is unknown about grain
interactions in powder flow, and in order to represent any sensitivity to
initial conditions. Methods from statistical inference
need to be used not only to fit the models, but also to assess the quality of the fitted models. Formal statistical methods will complement
fitting by eye and by intuition, since they will be able to describe and compare features of realizations that the eye cannot easily see. 

The methods outlined here are suited to studying two- or three-dimensional powder flow under conditions where the powder stays in a packed or close-to-packed state. Examples include flow in annular cells \cite{grebenkov}, flow in hoppers \cite{to:2005}, and flow during triaxial tests \cite{cheng:2004}. If the powder starts at rest, then the initial  arrangement of its grains in a model must be consistent with an arrangement which could arise from the physical process of specimen preparation. Once the powder is put into motion,
the joint trajectories of the grains in realizations of the model and
in replicates of the experiment must be
sufficiently similar that a reliable prediction can be made from the model.  Methods from statistical inference are required to assess whether or not there is evidence that the model is failing to capture the dynamics of the flow. When the models are deterministic and produce only a single prediction for any set of experimental conditions, no model validation procedure can demonstrate that a model represents the dynamics of powder flow. 

The word \textsl{statistic} is used to represent any number calculated
from observations of a physical flow process or its simulation. It
need not be an average, but could be a measure of variability around a
mean or an extreme value. The adjective \textsl{statistical} will be
used in reference to statistical inference, and not statistical
mechanics. The goal is to undertake model assessment by means of
hypothesis tests and other tools of inference which accommodate the
natural variability of the observations, rather than trying to
eliminate variability by means of a probabilistic limiting 
argument. Experiments will consist of many \textsl{replicates} conducted under
the same experimental conditions. For any given set of conditions, an indefinite number of
\textsl{realizations} can be generated from any fitted stochastic model of the
process.

\section{The Need for Stochastic Models}

Any powder flow phenomenon requires stochastic modeling. Given fixed experimental conditions, all replicates are subject to uncontrolled variation which cannot be eliminated by a probabilistic limiting argument. 

If a powder begins at rest, then it is generally in a disordered jammed state. This disordered state arises from a physical preparation process which is sensitive to initial conditions and which generates unpredictable arrangements of grains. It is possible to assume that the variability in structure between replicates has no effect on the subsequent flow behaviour, but this
would be a dangerous assumption to make when the powder
flows in a nearly-packed state.

If the physical phenomenon of interest depends strongly on the
initial contact network among the grains, then the contact network must
also be stochastically modelled. In many simulated packings and all data from
physical specimens, it is impossible to determine which grains are very close
and which grains are in contact \cite{aste}. Establishing a contact structure by
means of an arbitrary deterministic rule may result in the model
failing to sample from contact networks that are present in physical systems.

Once the powder is acted on by a force and flows, then the powder flow may be sensitive to initial conditions. This would require stochastic modeling even if all of the details of the grain interactions were known. Since the details of the grain interactions are unknown, modeling requires making choices about how to represent the many possible interactions among the grains. If stochastic elements are introduced into the grain interactions, then these elements can  represent the ignorance of interaction mechanisms better than an arbitrary deterministic model could.  

\section{Formulating Stochastic Models}

Any stochastic model for powder flow must sample from the ensemble of jammed initial configurations, the ensemble of viable contact networks for the sampled configuration, and the ensemble of joint trajectories that could emerge from the sampled contact network.  Ideally, these ensembles will be identical to those
that the physical process itself samples from when the process is
repeated many times. The physical process ensembles will be defined by a set experimental procedure, by the nature of the materials used, and
by certain macroscopic variables which govern the experimental
procedure and which are used to characterize acceptable outcomes. It
will not be possible to describe these ensembles by abstract
mathematical models, as is the case in conventional statistical
mechanics. 

In most cases, the model ensembles will differ from those of
the physical process. The usefulness of stochastic
models will depend on their being able to generate trajectories which cannot be distinguished from physical trajectories by means of statistical inference. Since there will be no way to
determine the form of a stochastic model from theory, it will be necessary
to try many different models and to select the best ones by
inferential methods. 

If a powder begins at rest, then a stochastic model for the packed
grains is required to represent the unpredictability of the physical process
which produced the initial packing. In the best circumstances, a large
number of specimens can be prepared by that physical process and
imaged by X-ray \cite{aste,aste2,richphil,seidler}, NMR \cite{sederman}, 
or confocal \cite{kohonen,toiya} methods. From these images, 
initial states can be selected at random directly from the physical 
ensemble for use in stochastic models of flow. 

If images of the internal structure of real specimens are not
available, then the packed states need to be generated by some type of
stochastic packing generator. There are no generators which
are known to sample from the same ensemble as a physical process. The
first generators used, based on 
ballistic methods \cite{vb:1972,aparacio,mueller:2005,coelho} 
or on the rearrangements of random point patterns
\cite{jt:1981,jt:1985,barg:1991,lubstil:1991,speedy:1998,zinchenko},
have no basis in the physics of packing formation at all. Models
based on Discrete Element Methods [DEM] 
\cite{cundall,yen,zhu,zhu:2008} are inspired by the physics of
packing formation, but all are based on unverifiable assumptions about
how grains interact. It is possible to avoid
these issues by assuming that any packing generator which can produce
the right mean volume fraction of grains is good enough, but
this is a dangerous assumption. If the generator
produces initial arrangements of grains which possess structure not found in
physical specimens, then any model of an aspect of powder flow which depends on initial grain arrangements may be useless.

Modeling of the contact network will require the use of an algorithm
which randomly chooses whether or not closely neighboring spheres are
in contact. These algorithms must avoid
selecting contact networks which are mechanically unstable. 

The modeling of powder flow is generally undertaken using DEM
models \cite{zhu:2008}. If the DEM model is not stochastic, then it will produce only
one predicted joint trajectory for any one initial arrangement of grains. This joint trajectory cannot be expected to belong to the ensemble of joint trajectories of the 
physical process if any of the assumptions about grain interactions
are incorrect. Even
if the single prediction is in the ensemble of trajectories for the
physical system, there is no way to tell if it is a typical trajectory
for that ensemble.  

Stochastic models for powder flow can be developed from DEM
models. To represent sensitivity to initial conditions, small random
perturbations of particle positions could be introduced. To represent ignorance
of the mechanisms of grain interaction, the model could randomly choose between
several models for grain interactions at each time step. This random selection might
be based on local conditions, and may involve random selection of friction coefficient 
values. A well-fitting stochastic model would sample from
an ensemble of trajectories which would approximate trajectories from the physical
ensemble to the extent that no method of statistical inference
could distinguish a sample of model trajectories from a sample of
physical trajectories. 

The construction of any models for initial packings, contact networks,
and powder flows should be consistent with what is known (as opposed
to what is assumed) about the physics of powder flows. Since the
models are stochastic, it is possible that models which are much simpler
than the physical process may be useful at representing some aspect of powder flow. When such models are proposed for reasons of
computational expediency, it is necessary to show by means of
statistical inference that these simplifications have no effect on the
usefulness of the model.

Given the complexity of powder physics and the arbitrary elements of
all models for powder flow, it is not reasonable to expect that a
single proposed model will faithfully capture all of the mechanical
properties of an arbitrary powder. Instead, it would be best to focus
on fitting models for one physical phenomenon of interest and for one
type of powder at a time, and then seeking generalizations once
well-fitting models are identified. Once the physical phenomenon is chosen, it is necessary to
summarize that phenomenon with a set of response statistics. These
statistics are calculated from realizations of each fitted model, and
describe whether or not the phenomenon occurred, or describe the
phenomenon. There should be as few response statistics as possible, so
as to make model fitting as simple as possible. The distributions of
the response statistics will be needed to create the intervals needed
for prediction. These distributions can only be studied through
data from multiple realizations of the model, since there is no theory
available to predict their form. 

\section{Fitting stochastic models}

In any stochastic model, the parameters can be classified as physical
or calibrational. Physical parameters are values such as the
acceleration of gravity, which are assumed to be universal and are
supplied by values from other experiments. The calibrational
parameters are the parameters of the random elements of the
model. Calibrational parameters can be further subdivided into those
which are associated with sensitivity to initial conditions and those
which are related to aspects of the model which reflect ignorance of
physical mechanisms. Both types of calibrational parameter must be
fitted by statistical means. There 
must be enough response information available from both the model and
the experiment in order to fit all of the parameters.

The primary basis of fit for a stochastic model is the relationship
between the distribution of the response for the model and the response for the physical experiments. The fit should be based on matching the mean response. The presence of
stochastic elements which represent the inability to model the details
of grain interactions may introduce extra variability into
realizations from the model. These stochastic elements may also change subtler aspects
of the distributions of the response from the model and from the
data. 

Comparing the distributions of the response for the model and the
experiment requires a sample of realizations from the model and a sample of  replications of the experiment. The number of realizations required to
compare means is relatively small, but many more are required to usefully
estimate variances, covariances, and subtler aspects of the response 
distributions. When small samples are used, there is a greater risk
that the model will be fit to idiosyncrasies of the particular 
experimental replicates instead of being fit to the underlying physics of the flow.

Objective fitting of the calibration parameters requires solving 
an optimization problem using numerical methods. If the response 
were a stress-strain curve from a triaxial test, then the mean of all
curves from the experimental replicates and the mean of all curves 
from realizations of the model could be found. Calibration parameters
could then be chosen to minimize the mean square difference between 
the two average curves, or to minimize some other measure of difference. The 
parameters could also be fit by means of expert guessing. This method 
may produce a useful set of parameters, but better solutions may be overlooked. 

\section{Assessing the fit of stochastic models} 

Once the model has been fit, the
quality of that fit must be assessed. If the fitted model is subject
to no further objective assessment, there is a great risk of fitting a
model which fails to represent the underlying
physics of powder flow. 

The subset of space occupied by the grains of a flowing powder at any point in time can be thought of as a realization of a random set \cite{math:1975,mol:2005}. These sets are random because their structure is unpredictable between
different replicates or realizations. Their internal structure is disordered,
but the disordered pattern of grains is neither stationary nor ergodic. 
There is no simple or obvious way to coordinatize the internal structure of an ensemble of 
disordered patterns, and so these patterns must be summarized
by sets of descriptive statistics. These statistics may be based on 
single patterns observed at a fixed time in each realization or replicate, 
or they could be constructed from many patterns observed at fixed
times along the joint trajectory of the grains.

To be useful, a set of descriptive statistics has to be able to identify
common aspects of all realizations from the experiment and to identify
any systematic differences between the joint trajectories of model 
realizations and experimental replicates. Finding these statistics is
challenging, since 
spatial statistics often lack the power to clearly distinguish outcomes 
from different spatial processes \cite{badsil}. Using the random set 
analogues of the mean and covariance alone will also not suffice, since
random set processes are highly non-Gaussian. If model verification
is based on matching one or two spatial statistics between the model 
and the experiment, there is a risk that these statistics will agree 
because they lack the statistical power to distinguish the two processes,
and not because the model in any way represents the physics of 
powder flow. 

To assess the fit of a fitted model, it is necessary to compare the realizations 
and the replicates using many different statistics, all of which summarize
different aspects of the disordered patterns within each realization or 
trajectory. Some of these statistics will be chosen to seek out possible
differences between model and experimental trajectories based on
what is known about the physics of powders, but others must be chosen 
from a large library of descriptive statistics which compare other aspects
of the trajectories. These additional statistics are required to seek out
differences which neither the eye can see nor existing theory 
would suggest looking for. It will be necessary to develop extensive
libraries of statistics for this purpose, which will include $k-$point correlation functions, point process statistics \cite{stoy:1995},
tessellation-based statistics \cite{finney,aste}, statistics based on mathematical models 
for physical processes applied in a non-physical context \cite{picka:2005},
statistics found useful in the analyses of other experiments, and many other
descriptors of ordered structure which have not yet been invented. 

Since the models idealize the physical processes, statistics will be found which 
can identify differences between realizations of the model and experimental replicates. These differences will be irrelevant if those 
statistics do not affect the response. If it were possible to classify
beforehand exactly which statistics affected the response, then the 
assessment could be based on those statistics alone. Since little is 
known about the relationships between particular responses and
other descriptive statistics, these relationships also need to be established 
by methods from statistical inference.  The statistical analysis 
should be based on replicates from physical experiments. If data from 
experiments are not available, then relationships could be sought from realizations of the
model. Using the model is risky, as relationships identified from a flawed model might differ from those found from the physical data. 

The assessment of model fitness is a problem in multivariate statistical 
inference \cite{johnwich}. While it is impossible to establish if the joint distributions of the
descriptive statistics for model and experiment are the same, three different 
methods can be used to look for evidence that the distributions are different.
All methods work best when the number of observations is large and the 
number of statistics is small. If there are many statistics and few observations,
then no method will produce trustworthy results.  

Multivariate distributions can be compared by means of statistical tests which 
seek to identify differences in ensemble means. These tests are generally based on the
assumption that both joint distributions are Gaussian, which may not
be true. If a very small number of observations are all that can be
found from the experiment and a joint distribution can be fit to
statistics from many realizations of the model, then it is also
possible to test how unlikely it is that the experimental observations
came from the model distribution. This type of test is risky, since
the observations in a small experimental sample may be unrepresentative. A third method of comparison is based on statistical
learning, also known as data mining \cite{htf}. A classification rule is
constructed with the aim of being able to distinguish replicates  
of the experiment from realizations from the model. If a subset of variables can be
found which can be used to build an effective classification rule,
then these variables may identify differences between the model and the experiment. 

The fit of the model may also be assessed by eye. If a simulated movie
generated by the model is visually indistinguishable from a movie of a
physical experiment, then it may be tempting to assume that the model
fits. This approach assumes that the vision of an expert can identify
the mean differences between realizations of the model and replicates
of the experiment, and can identify which of these differences affects
the response. It is possible to construct
examples of spatial processes whose realizations cannot be
distinguished by eye, yet can be distinguished by statistical methods. 
If an expert does claim to see differences between model and
experimental output, then the expert may not be able to express the
basis of their claim or to prove that their claim is free of any conscious or
unconscious bias.  

No final and absolute decision on the quality of a fitted model can be
made from the outcome of a single experiment. The assessments are
based on statistical procedures, and so may be subject to errors. If
the replicates of the experiment do not form a representative sample
from the ensemble, then they could provide misleading evidence against
an otherwise useful model. The risk of this happening is 
high if the number of replications is small. Alternatively, no
evidence may be found against the model because no statistic has yet
been found which can identify important differences between the model
and the experiment. The probabilities of these errors can only be reduced through
using larger numbers of replications within experiments and through
being ingenious in the development of new statistics. The emergence of a physical explanation 
for the response may only occur after comparison of fitted models arising
from repetitions of the experiment at many different sites 
using the same experimental protocols. 

\section{Conclusions}

The objective validation of models of powder dynamics is a problem in 
spatial statistical inference. It requires that the models be stochastic, to reflect
both ignorance of the details of grain interactions and any sensitivity to
initial conditions. These models will not exactly reproduce any one
observed trajectory, but will be useful if they can produce responses 
with the same distribution as an experiment and if a thorough fitness
assessment reveals no physically significant differences between trajectories 
from  the model and from  the experiment. 

Implementation of objective validation requires significant work by 
scientists and statisticians. Scientists must devise the experiments,
quantify the physical property of interest with response statistics, and 
develop accurate methods of imaging the full three-dimensional structure of a powder 
in motion. Statisticians need to develop useful methods for comparing the
joint distributions of large numbers of descriptive statistics, and to determine
how much information is required for these methods to be effective. Statisticians
and scientists must jointly develop new descriptive spatial statistics, study their properties, 
and index them in libraries that can be used for future analyses. General 
methods for modeling powders can only emerge after common elements 
from the analyses of many experiments are identified.

This approach to model validation and development mimics the approach
taken in the development of the first thermodynamic models. It is
based on experimentation and objective validation of model fit,
and not on extrapolation of previously existing and successful models derived
for simpler phenomena. If the models developed can be thought of as a
thermodynamics of powders, then they differ from classical models in
having random state variables, in being intended for history-dependent processes, 
and in accommodating the intense
multiparticle interactions which dominate dense powder flow. 
If an inference-based approach to modeling is not taken, there is a
significant risk that the fitting of models could only compress data from specific experiments, rather than the summarizing 
the physics of powder flow. 


\end{document}